\documentclass[aps,prl,reprint,superscriptaddress]{revtex4-1}
\usepackage{times}
\usepackage{graphicx}
\usepackage{hyperref}

\usepackage{xcolor}

\begin{document} 

\title{Loss and decoherence at the quantum Hall - superconductor interface}

\author{Lingfei Zhao}
\email{lz117@duke.edu}
\affiliation{Department of Physics, Duke University, Durham, NC 27708, USA}

\author{Zubair Iftikhar}
\affiliation{Department of Physics, Duke University, Durham, NC 27708, USA}

\author{Trevyn F.Q. Larson}
\affiliation{Department of Physics, Duke University, Durham, NC 27708, USA}

\author{Ethan G. Arnault}
\affiliation{Department of Physics, Duke University, Durham, NC 27708, USA}

\author{Kenji Watanabe}
\affiliation{National Institute for Materials Science, 1-1 Namiki, Tsukuba 305-0044, Japan}

\author{Takashi Taniguchi}
\affiliation{National Institute for Materials Science, 1-1 Namiki, Tsukuba 305-0044, Japan}

\author{Fran\c cois Amet}
\affiliation{Department of Physics and Astronomy, Appalachian State University, Boone, NC 28607, USA}

\author{Gleb Finkelstein}
\affiliation{Department of Physics, Duke University, Durham, NC 27708, USA}

\date{\today}


\begin{abstract} 
We perform a systematic study of Andreev conversion at the interface between a superconductor and graphene in the quantum Hall (QH) regime. We find that the probability of Andreev conversion from electrons to holes follows an unexpected but clear trend: the dependencies on temperature and magnetic field are nearly decoupled. We discuss these trends and the role of the superconducting vortices, whose normal cores could both absorb and dephase the individual electrons in a QH edge. Our study may pave the road to engineering future generation of hybrid devices for exploiting superconductivity proximity in chiral channels.  
\end{abstract}

\maketitle  

Combining superconductors and quantum Hall (QH) systems has been proposed as a particularly promising direction for creating novel topological states and excitations~\cite{Mong2014}. Over the past few years, significant progress has been achieved in developing such hybrid structures~\cite{Rickhaus2012,Komatsu2012,Wan2015,Amet2016,Lee2017,Park2017,Sahu2018,Matsuo2018,Kozuka2018,Seredinski2019,Zhao2020,Gl2022,Hatefipour2022}. In particular, hybridization of QH edge states across a narrow superconducting wire is expected to create a gapped topological superconductor~\cite{Clarke2014,Lee2017}. In the fractional QH systems, the strong interactions potentially fractionalize Majorana fermions into parafermions~\cite{Clarke2013,Gl2022}, a key ingredient for universal topological quantum computing~\cite{Mong2014} and exotic circuit elements such as fractional charge transistors~\cite{Clarke2014}. 

At the interface between a superconductor and a QH system, the QH edge states are expected to be proximitized, turning into chiral Andreev edge states (CAES). These are dispersive states which hybridize the electron and the hole amplitudes~\cite{Hoppe2000}. An electron approaching the superconducting region is converted to a linear combination of CAES, which interfere as they propagate along the interface. The outgoing particle can either stay as an electron or turn into a hole. We have previously observed clear evidence of the electron-hole conversion in the quantum Hall devices with superconducting contacts~\cite{Zhao2020}.
However, the exact mechanism of this conversion remains open: the role of the disorder, superconducting vortices, and the exact nature of the CAES in this system have been discussed. It is known that in order to observe a strong Andreev conversion in an ideal system a precise matching between the superconductor and the QH edge state momenta is required~\cite{Giazotto2005,Akhmerov2007,vanOstaay2011,Michelsen_supercurrent_2022}. However, the presence of disorder is expected to relax this constraint~\cite{Manesco_mechanisms_2021,Kurilovich_disorder_2022}. Vortices in the superconductor and extra QH channels induced by doping could also modify the signal~\cite{Kurilovich_disorder_2022,Tang_2022,Manesco_mechanisms_2021,David_effects_2022}. In particular, the normal cores of the vortices can absorb the electrons and holes, reducing the amplitude of the measured signal~\cite{Zhao2020,Kurilovich_disorder_2022,Schiller_interplay_2022}.  

To address the microscopic mechanisms affecting the Andreev conversion, here we perform a systematic study of the conversion probability vs temperature, $T$, magnetic field, $B$, and interfacial length, $L$. We find that the dependence of the electron-hole conversion probability on magnetic field and temperature nearly factorizes. We suggest a simple phenomenological expression involving exponential decays as a function of $B$, $T$ and $L$, and a prefactor determined by the configuration of superconducting vortices. The expression captures the observed dependencies very well, and can be interpreted in terms of the CAES loss and decoherence. We finally discuss the distribution of the Andreev conversion probability~\cite{Kurilovich_disorder_2022}, which is unexpectedly found to have a roughly triangular shape. We discuss the implications of our findings for the future development of the more complex devices, which will further explore the physics of superconducting correlations in the chiral states. 

\begin{figure}
\includegraphics[width=1\columnwidth]{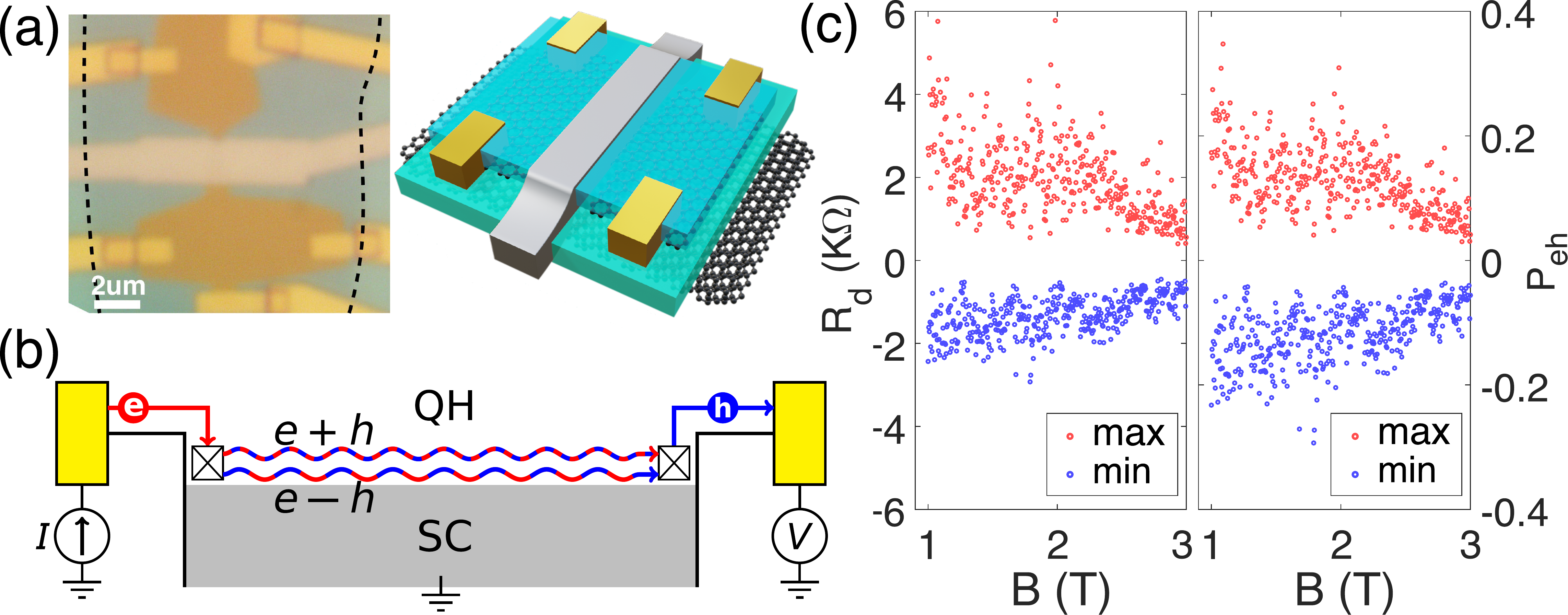}
\caption{(a) Optical image (left) and 3D schematics (right) of the device. The black dashed lines label the boundaries of the graphite gate underneath. The yellow electrodes are normal Cr/Au contacts and the light gray is the superconducting MoRe, which forms 0.5 $\mu$m and 1 $\mu$m interfaces with graphene. (b) Sketch of the measurement setup. The current is injected from the upstream contact into the grounded superconductor, while measuring the voltage at the downstream contact. (c) The maximum and minimum values of $R_d$ (left) and $P_{eh}$ (right) measured for the shorter interface. {The statistical information is collected in the  $V_G$ range corresponding to $\nu= 2$, as outlined by the green lines in Fig. 3a.} The temperature is 40 mK. }
\end{figure}

The main device studied here is a hBN/graphene/hBN heterostructure in contact with both superconducting and normal electrodes (Fig.~1a). The superconducting electrode (light gray) is made of sputtered Mo-Re alloy (50-50 in weight) with a critical temperature $T_c\sim\ $10 K and an upper critical field $H_{c2}$ exceeding 12 T. The work function of the alloy $\sim\ $4.2 eV~\cite{Haas2008} is slightly lower than that of graphene $\sim\ $4.5 eV~\cite{Liang2015}, resulting in an n-doped graphene region nearby.  The graphene sheet is separated into two independent regions by the etching step that defines the superconducting electrode. The widths of the resulting superconductor-graphene interfaces are 0.5 and 1 $\mu$m. The normal contacts (yellow) are thermally evaporated Cr/Au. The electron density inside the graphene is controlled by applying a voltage $V_{G}$ to the {graphite gate which spans the whole area underneath the heterostructure. While such graphite gates are known to efficiently screen the disorder potential, the results here are similar to our previous measurements of samples without the graphite gate. This indicates that the observed physics is not strongly influenced by the disorder in the graphene layer.}

{We measure the nonlocal resistance downstream of a grounded superconducting contact, $R_d=dV_d/dI$, as sketched in Fig.~1b and  demonstrated in the supplementary~\cite{supp}.  The CAES formed by the superconductor travel along the interface and recombine into either an electron or hole (or their linear combination) at the end of the interface. Sweeping the gate voltage $V_{G}$ on top of the QH plateau tunes the momentum difference between the interfering CAES and produces an oscillating pattern of $R_d(V_G)$ (Fig.~S1 in \cite{supp})~\cite{Zhao2020}. Due to the disordered nature of the interface, these interference patterns are highly irregular and resemble the universal conductance fluctuations~\cite{Manesco_mechanisms_2021,Kurilovich_disorder_2022}. 
Throughout the paper we analyze the statistical properties of these patterns.}

{We further convert $R_d$ into the difference between the probabilities of normal and Andreev reflections, $P_{eh} \equiv P_e-P_h$, where $P_e$ ($P_h$) is the probability of an electron (or a hole) to be emitted downstream of the superconductor. 
It is straightforward to show that $P_{eh}=R_d/(R_d+R_H)$, where $R_H$ is the Hall resistance~\cite{Zhao2020}. Note that the extreme values of $R_d$ are reached either for the pure electron reflection ($P_e=1$, $P_h=0$ and $R_d=\infty$, the interface is effectively fully opaque), or for perfect Andreev conversion ($P_e=0$, $P_h=1$, $R_d=-R_H/2$, a Cooper pair is transferred across the interface per incoming electron). We therefore expect that the distribution of $R_d$ should be skewed toward positive resistances.} 

{Indeed, this skewness  can be observed by studying an imbalance between the maximum and minimum values of the downstream resistance $R_d (V_G)$. We extract these quantities in the $V_G$ range corresponding to the $\nu=2$  plateau for a given field and then plot $R_{max}$ and $R_{min}$ as a function of $B$ in Fig.~1c. The field changes the vortex configuration every few mT~\cite{Zhao2020}, thereby allowing us to sample multiple different patterns of $R_d(V_G)$. The data clearly shows that $R_{max}$ is on average larger than $|R_{min}|$. The apparent imbalance between electrons and holes is eliminated by converting $R_d$ to $P_{eh}$ in Fig.~1c. (See Fig.~{S3} in \cite{supp} for the similar result measured in another device.) We conclude that the probabilities of an electron or a hole being emitted downstream ($P_e$ and $P_h$) are very similar.  From now on, we present $P_{eh}$ in lieu of $R_d$. }

\begin{figure}
\includegraphics[width=1\columnwidth]{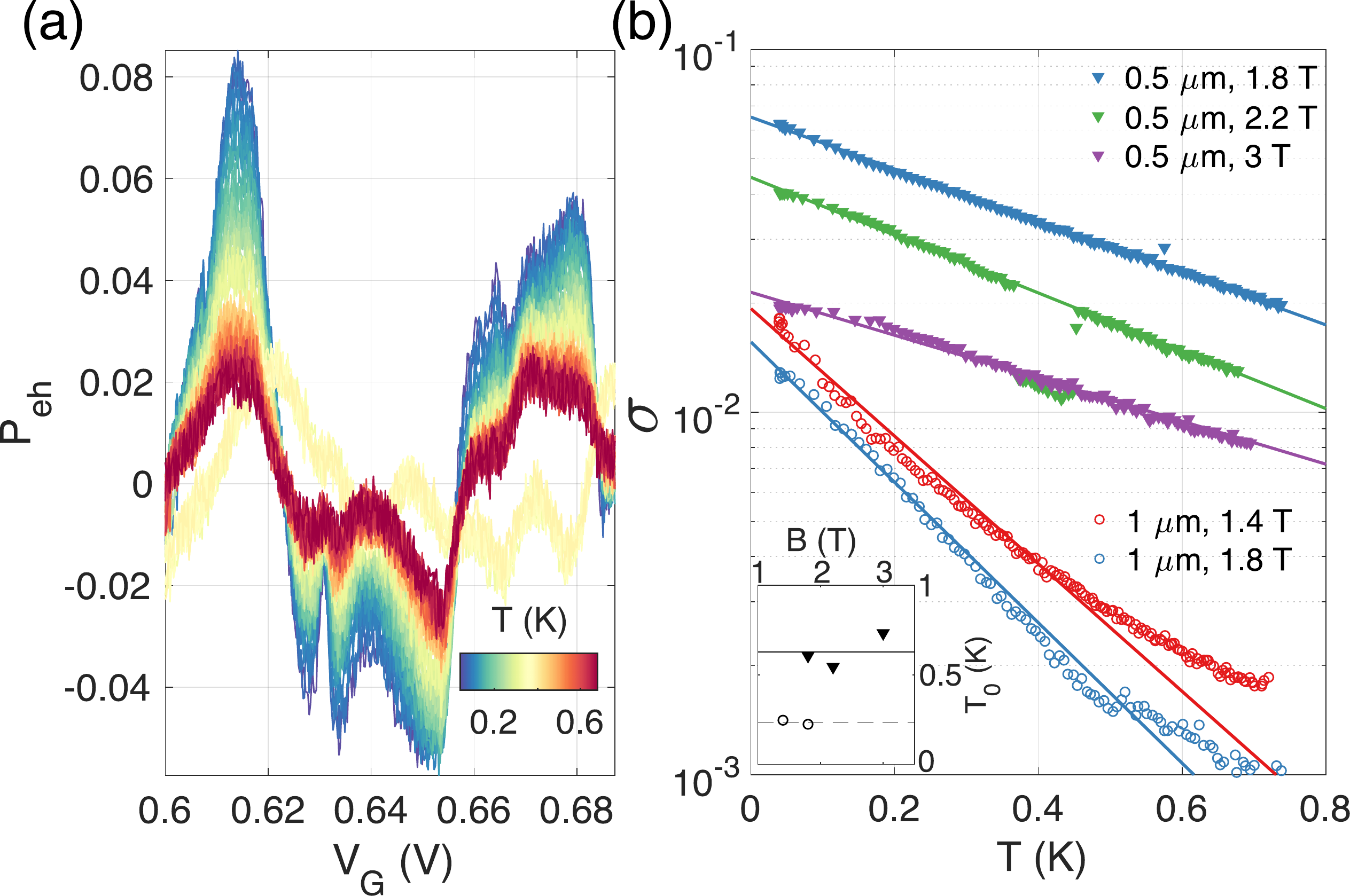}
\caption{(a) $P_{eh}$ measured for the 0.5 $\mu$m interface as a function of $V_G$ on top of the $\nu=$ 2 plateau at $B=$ 2.2 T. The temperature is varied from the base temperature of 40 mK to about 670 mK. (b) The standard deviation of $P_{eh}$ of the 0.5 $\mu$m (filled triangle) and 1 $\mu$m (open circle) interfaces as a function of temperature at various magnetic fields. The inset plots the decay constant $T_0$ obtained from exponential fits vs. the magnetic field. The lines represent the averages of the dots.}
\end{figure}
 
{Depending on the length of the interface, the highest electron-hole conversion efficiency we observe is about 0.2 -- 0.3. While being very high, this value is still far from reaching unity. Thermal smearing, decoherence and tunneling into the normal vortex cores can all contribute to this suppression. Since at zero temperature only the effect of the vortices remains, we first look into the dependence of $P_{eh}$ on temperature.} Fig.~2a plots $P_{eh}$ of the shorter interface (0.5 $\mu$m) measured over the $\nu=$ 2 plateau at $B=$ 2.2 T from 40 mK to 670 mK. As the temperature $T$ increases, $P_{eh}$ gradually decays towards zero, except around 400 mK where the traces briefly jump to a completely different oscillation pattern. We attribute this jump to a temporary change of the configuration of superconducting vortices during the measurement. 

To get a quantitative understanding of the thermal effects, we plot the standard deviation of the traces, $\sigma$, as a function of temperature in Fig.~2b for both interfaces at various magnetic fields. The filled triangles (open circles) represent $L=$ 0.5 $\mu$m (1 $\mu$m). The green triangles correspond to the data in Fig.~2a. We find an exponential decay of $\sigma$ as a function of temperature, i.e. $\sigma\propto \exp(-T/T_0)$, where $T_0$ is the decay constant. Remarkably, $\sigma$ follows roughly the same decay rate for a given interface length. Even the curves of Fig.~2a which experienced a random jump follow the same slope. (Note the few green symbols nearly overlapping with the purple ones around 0.4 K.) This means that the configuration of superconducting vortices does not have a strong influence on the temperature dependence, even if it dramatically affects the amplitude and the pattern of fluctuations! 

The exponential decay is observed regardless of the length of the interface and the magnetic field. For the longer interface, the decay rate becomes less steep at higher temperatures, but only as the signals approach the noise floor of a few $\times 10^{-3}$, so we extract the slope from the low-temperature range. In the inset of Fig.~2b, we plot the resulting $T_0$ vs. $B$ for $L=$ 0.5 and 1 $\mu$m. It is clear that $T_0$ does not show any strong dependence on $B$ and scales inversely with $L$. 

In principle, the exponential temperature dependence could originate from the averaging over energy, like the thermal smearing of oscillations in a QH Fabry-Perot interferometer. However, the effects of thermal smearing are expected to saturate below $T_0$, while we observe an exponential decay down to temperatures more than 10 times lower (for the shorter interface). Moreover, thermal smearing is predicted to be rather inefficient in this system~\cite{Kurilovich_disorder_2022}, because Andreev conversion is only weakly energy dependent. Furthermore, thermal broadening would smear the fluctuations of $P_{eh}$ over $V_G$. Instead, we observe
that the fluctuations in Fig.~2a uniformly decay with temperature with no noticeable smearing. In fact, we show that curves in Fig.~2a could be rescaled to nearly match (see Fig.~{S5} in \cite{supp}). 

We therefore ascribe the temperature decay of $P_{eh}$ to the decoherence of the CAES. Indeed, decoherence resulting in the exponential suppression of the oscillations $\propto \exp(-T/T_0)$ has been measured in Mach-Zehnder interferometers~\cite{Ji2003,Roulleau2008}. However, the coherence length of the QH interferometers usually exceeds the length of the superconducting contact by more than an order of magnitude, including our own measurements using similar graphene samples~\cite{Zhao_graphene_2022}. The rapid decoherence of the CAES as compared to the QH edge states is likely due to their hybrid nature: The electron and the hole components should acquire an opposite phase in the fluctuating electrostatic potential. The source of these fluctuations could be the gate, or the vortex cores, which serve as a reservoir of normal electrons located next to the CAES.

Since $T_0$ is not strongly affected by the magnetic field and the jumps of the vortex configuration, the effects of the vortices should be primarily encoded in the zero temperature value of $\sigma$. The base temperature of 40 mK $\ll T_0$ allows us to use the corresponding values of $\sigma$ as a close approximation to zero temperature value. In the following section, we study the dependence of this quantity on the magnetic field.

\begin{figure}
\includegraphics[width=1\columnwidth]{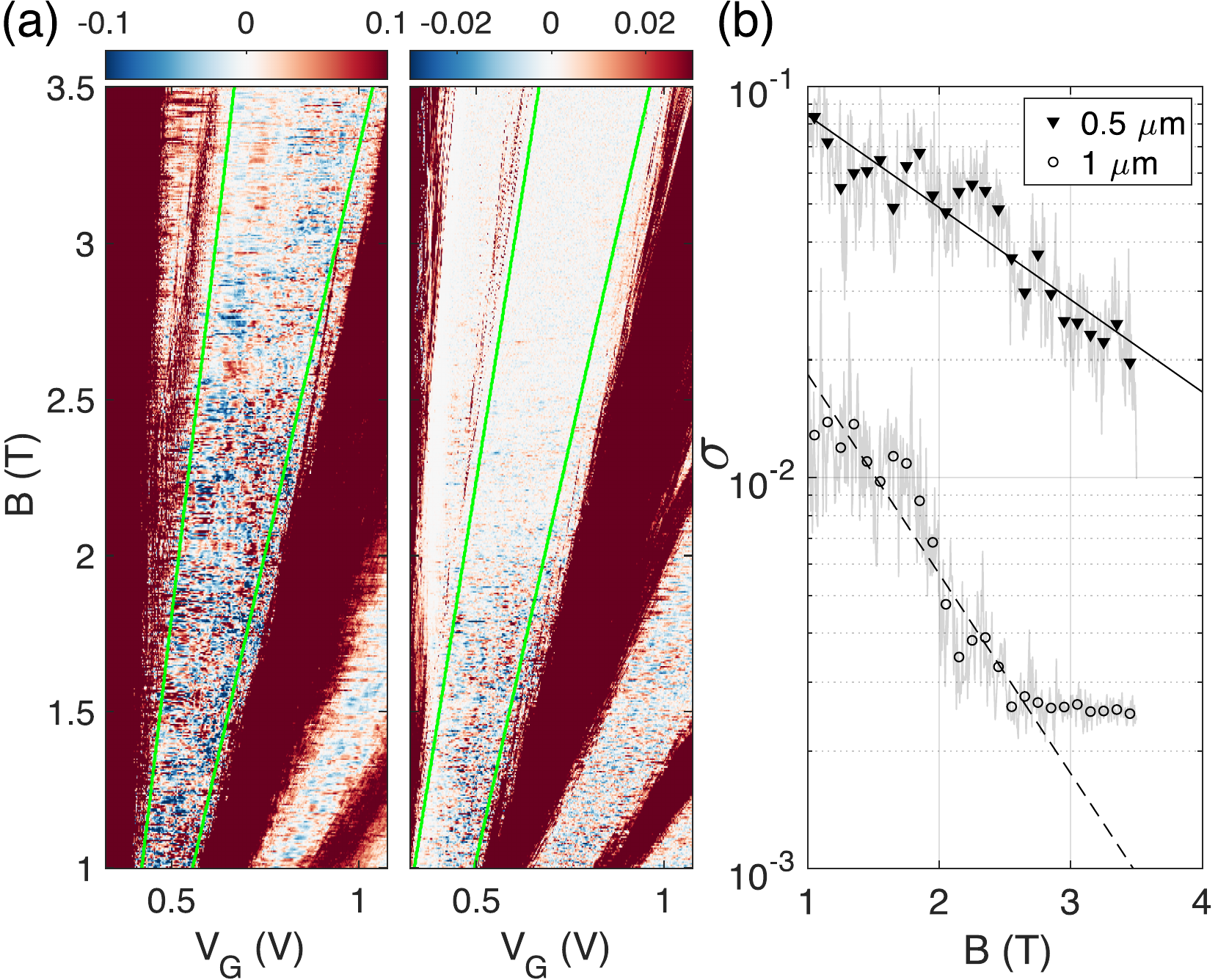}
\caption{(a) $P_{eh}$ plotted against $B$ and $V_G$ for $L=$ 0.5 $\mu$m (left) and 1 $\mu$m (right) at $T=40$ mK. The green lines mark the region of the $\nu=$ 2 plateau used for the analysis in panel (b) and Fig.~1c. 
(b) {The standard deviation $\sigma$ of $P_{eh}$ measured over the $V_G$ range corresponding to $\nu=2$. The top and bottom gray curves correspond to the $L=$ 0.5 and 1 $\mu$m interfaces. The lower curve saturates upon reaching the noise floor of the experiment, at $\sigma \approx 2.5 \times 10^{-3}$. The symbols (circles and triangles) represent $\sigma$ averaged over 100 mT range, $\left<\sigma\right>_{\Delta B}$. This averaging reveals the overall decay trend of $\sigma$, represented by the exponential fits (straight solid and dashed lines).}}
\end{figure}

In Fig.~3a we show the patterns of $P_{eh}$ over a wide range of magnetic field for the 0.5 $\mu$m interface (left) and the 1 $\mu$m interface (right). A corresponding Hall conductance map is shown in Fig.~{S2}~\cite{supp}. The boundaries of the $v=$ 2 plateau are labeled by the green lines. The amplitude of $P_{eh}$ for the longer interface is not only smaller but also decays much faster with increasing $B$. This is consistent with our picture of electrons/holes tunneling from the edge state into the normal cores of the superconducting vortices. Indeed, the probability grows with both $L$ and the vortex density, which increases with $B$. 

While the dependence of $P_{eh}$ on $B$ is highly stochastic (Fig.~1c \& {S6} in \cite{supp}), we can analyze the large-scale trend by averaging $\sigma$ over a relatively small range of magnetic fields, $\Delta B=100$ mT. The resulting quantity averages over multiple vortex configurations and will be denoted by $\left<\sigma\right>_{\Delta B}$. As shown in Fig.~3b, it roughly follows an exponential decay with $B$ for both interfaces. Excluding the points that reach the noise floor, an exponential fit produces $B_0=$ 1.84 T for $L=$ 0.5 $\mu$m and $B_0=$ 0.85 T for $L=$ 1 $\mu$m, in agreement with the expected relation $B_0\propto1/L$. 

Note that this relation further implies an exponential decay of $\left<\sigma\right>_{\Delta B}$ as a function of $L$, $\left<\sigma\right>_{\Delta B}\propto\exp\left(-L/L_0(B)\right)$ with $L_0\propto1/B$ at zero temperature. This dependence indicates that the electrons/holes are efficiently absorbed by multiple vortices, so that the inverse decay length $1/L_0$ is proportional to the vortex density $\propto B$~\cite{Kurilovich_disorder_2022}. The $\exp\left(-L/L_0\right)$ dependence is directly confirmed in another device with three contacts of different interface length, as shown in Fig.~{S4}~\cite{supp}. Interestingly, both devices have similar $L_0$ in the $250-300$ nm range at 1.5 T. 

{To summarize the observed trends, we can write \begin{equation}
\sigma=A(\{\textbf{r}_{vortex}\})\exp\left(-\frac{B}{B_0(L)}\right) \times \exp\left(-\frac{T}{T_0(L)}\right)
\label{factorization}
\end{equation}
where prefactor $A$ represents fluctuations on the scale of a few mT due to vortex rearrangements.
Importantly the decay rate $B_0$ is mostly independent of $T$, and $T_0$ is mostly independent of $B$ (see Fig.~2b), which supports our assertion that different mechanisms are responsible for the $B$ and $T$ dependencies.} 

\begin{figure}
\includegraphics[width=1\columnwidth]{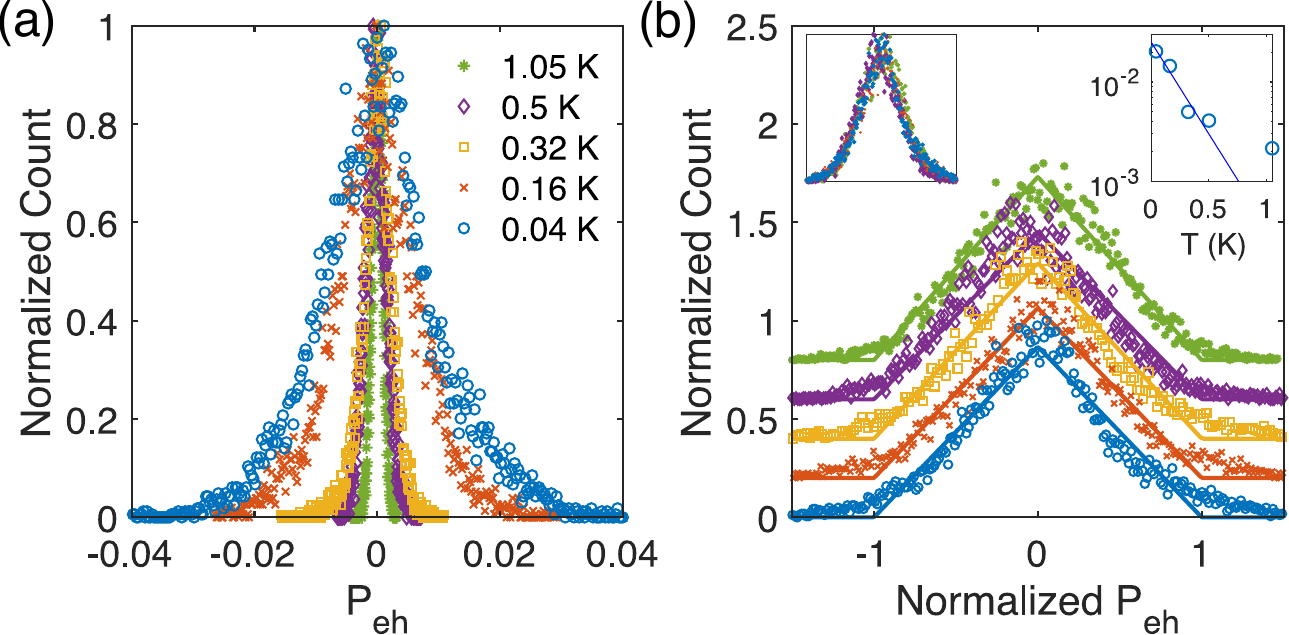}
\caption{(a) Histograms of $P_{eh}$ collected for the longer interface in the $1.4-1.5$ T window on the $\nu=$ 2 plateau at various temperatures. (b) The same data as in (a) rescaled by fitting a {phenomenological triangular shape (lines) and normalizing x-axis by the fitted width of the triangle. The curves are offset by 0.2 along the y-axis. Left inset: the same normalized distributions replotted without an offset. Right inset: the fitted width of the triangle plotted vs. $T$. The line is a guide for the eye, which follows the decay constant $T_0$ obtained in Fig.~2b.}}
\end{figure}

{So far we have only considered the standard deviation of $P_{eh}$, but it is informative to further look into its full distribution. In Fig.~4a we plot the histogram of $P_{eh}$ of the $1 \mu$m interface over a small range of $B= 1.4-1.5$ T. The range $\Delta B = 100$ mT is selected because on the one hand it covers many vortex rearrangements; on the other hand $\Delta B = 100$ mT $\ll B_0= 0.85$ T so that the systematic reduction of the amplitude $\exp(-\Delta B/B_0)$ is negligible. The measured histogram is clearly peaked around zero, which contrasts with the uniform (rectangular) distribution predicted in Ref.~\cite{Kurilovich_disorder_2022}.} 

{While the distribution gets narrower with increasing temperature, the shape remains roughly the same, as seen in Fig.~4b, which replots the histograms with the x-axis normalized by their widths. These widths are plotted in the right inset of Fig.~4b (circles), together with an exponential line whose slope is determined by the constant $T_0$ obtained from Fig.~2b. The decay of the distribution widths with temperature agrees with the line, except for the highest temperature, which is likely limited by the noise in determining $P_{eh}$. }

The distribution shape is expected to be influenced by the positions of the vortices, resulting in the random spread of $A\left(\{\textbf{r}_{vortex}\}\right)$ (see Fig.~1c \& {S6} in \cite{supp}). To evaluate and eventually eliminate the variations of the $P_{eh}$ due to vortex rearrangements, we normalize each $P_{eh}(V_G)$ trace by its standard deviation calculated for the same $B$. The resulting histograms of $P_{eh}(V_G)/ \sigma$ are shown in Fig.~{S7}~\cite{supp} and appear to have a roughly triangular shape similar to those in Fig.~4. We conclude that vortex rearrangements, while quite noticeable, cannot explain the shape of the $P_{eh}$ distribution. 

{In principle, the non-rectangular distribution of $P_{eh}$ may point to factors that have not yet been fully understood: spin-orbit coupling, interface transparency, contact doping, or specifics of graphene band structure. 
In particular, if the Andreev conversion probabilities for the two spins states are sufficiently different, the convolution of their rectangular distributions would produce a triangle. Regardless of these mechanisms, the distribution of the $P_{eh}$ should be strongly shaped by the particle losses.
In the supplementary~\cite{supp}, we present a toy model which qualitatively reproduces the observed distribution (Fig.~S8). In the model, the CAES propagation along the interface is represented by the evolution of a $2 \times 2$ electron/hole density matrix, represented by a trajectory on the Bloch sphere~\cite{Kurilovich_disorder_2022}. The key insight of the model is that the losses have uneven effect on different trajectories: Most of the trajectories undergo heavy losses, resulting in a small $P_{eh}$ that contributes to the center of the distribution peak. The rare trajectories which suffered less losses contribute to the tails of the distribution.} 

In summary, we have reported a systematic investigation of the Andreev conversion probability for the CAES states propagating along a superconducting contact in the QH regime. {The main result is the near decoupling of the $P_{eh}$ dependencies on the temperature and magnetic field, i.e. 
Eq.~\ref{factorization}. The dependence on magnetic field, $A(\{\textbf{r}_{vortex}\})\exp\left(-B/B_0\right)$ describes the particle loss, where the characteristic field $B_0$ decays with the interface length $L$ in agreement with the theoretical prediction~\cite{Kurilovich_disorder_2022}.} It is natural to associate the temperature dependence, $\exp\left(-T/T_0\right)$, with decoherence, which is found to be much more efficient compared to the conventional edge states in quantum Hall interferometers. Indeed, the relative phase of the electron and hole components of the CAES should be strongly influenced by the fluctuations of electric environment. Understanding and controling the loss and decoherence processes is crucial for the development of future devices that utilize superconducting correlations in chiral states.

\begin{acknowledgments}

We greatly appreciate stimulating discussion with A. Akhmerov, H. Baranger, L. Glazman, V. Kurilovich, A. Manesco, Y. Oreg, E. Sela, and A. Stern. 
Sample fabrication and characterization by L.Z. and E.A. were supported by NSF award DMR-2004870. Transport measurements by L.Z. and T.L. and data analysis by L.Z., Z.I. and G.F. were supported by the Division of Materials Sciences and Engineering, Office of Basic Energy Sciences, U.S. Department of Energy, under Award No. DE-SC0002765. The deposition of MoRe performed by F.A. was supported by a URC grant at Appalachian State University.
K.W. and T.T. acknowledge support from the Elemental Strategy Initiative conducted by the MEXT, Japan, (grant no. JPMXP0112101001), JSPS KAKENHI (grant no. JP20H00354) and CREST (no. JPMJCR15F3, JST).
The sample fabrication was performed in part at the Duke University Shared Materials Instrumentation Facility (SMIF), a member of the North Carolina Research Triangle Nanotechnology Network (RTNN), which is supported by the National Science Foundation (Grant ECCS-1542015) as part of the National Nanotechnology Coordinated Infrastructure (NNCI).

\end{acknowledgments}


%

\section*{Methods}

\subsection*{Device Fabrication}
The graphene and hBN flakes are exfoliated onto Si/SiO$_2$ substrates (280 nm thickness) with Scotch Magic tapes and then assembled together using PC/PDMS stamps. The thickness of the hBN flakes are selected to be $\sim$ 30 nm for top layers and $\sim$ 60 nm for bottom layers. The graphite back-gate layers are about 5 to 10 nm thick. The bottom hBN are intentionally chosen thicker to prevent breakdown between the electrodes and the graphite back-gate. The stamps are made by first dripping SYLGARD 184 silicone droplets onto hot glass slides ($170\sim180^\circ$C) to form PDMS hemispheres and then plasma ashing for 15 mins. We carefully place PC films over the PDMS bases to avoid bubbles in between. The PC films are prepared with PC solution (6\% in weight) spread between two glass slides. The stamps are first cured over a hotplate at 180$^\circ$C for 20 mins and then cured again over the stamping station upside down right before usage. During the dry transfer process, the ambient temperature is maintained above 90$^\circ$C to reduce water contamination. At the last stage, the fully assembled hBN/graphene/hBN/graphite heterostructure is slowly dropped onto a plasma ashed hot ($160\sim180^\circ$C)  Si/SiO$_2$ substrate to drag bubbles out. After removing PC residue in hot chloroform, the sample is annealed in open air at 500$^\circ$C over night.

The stack is characterized by AFM and Raman topography. We pattern the device over the identified clean region using e-beam lithography on top of double layer PMMA resist (495K, A4), developed in cold IPA/DI water mixture (3:1 in volume). Right before electrodes deposition, the patterned region is etched by CHF$_3$/O$_2$ (50/5 ccm, 70 W, 100 mtorr) carefully to etch through the top later hBN only. The normal metal electrodes are made of thermally evaporated Cr/Au (1/90 nm) with a base pressure 10$^{-7}$ torr. The superconducting electrodes are made of MoRe alloy (50-50 in weight) DC sputtered in a high vacuum system (AJA, 10$^{-8}$ torr).

\subsection*{Measurements}

The measurements are performed in an Oxford dilution refrigerator equipped with resistive coax lines together with two-stage low pass LC and RC filters. The total resistance of each filtered line is about 1.7 kOhm and the cut off frequency is around 200 kHz. The nonlocal differential resistances are measured using a 1nA square-wave excitation with home-made preamplifiers and USB-6363. 
The graphite back-gate voltage is supplied with USB-6363 in series with a 1/4 voltage divider. For consistency, and to reduce sample overheating, the magnetic field ramp direction is always from high to low.

\clearpage
\onecolumngrid

{\centering{\large \bf Supplemental Material}\par}

\global\long\def\theequation{S\arabic{equation}}
\global\long\def\thefigure{S\arabic{figure}}
\setcounter{equation}{0}
\setcounter{figure}{0}

\section*{Comparison with a non-superconducting interface}

\begin{figure*}[h!]
\centering
\includegraphics[width=1\textwidth]{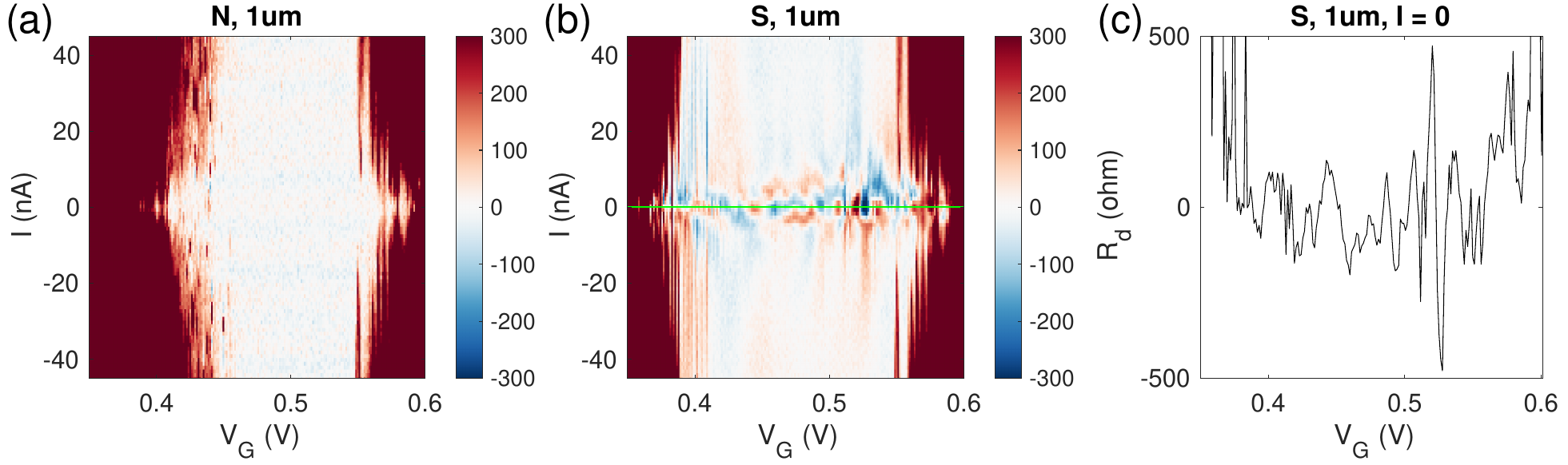}
\caption{Maps of the downstream resistance, $R_d$, plotted vs. bias $I$ and $V_G$ across $\nu=2$ plateau for the $1 \mu$m wide contacts: (a) normal and (b) superconducting. Downstream signal is equal to zero for the normal contact. For the superconductor, the zero-bias downstream signal is shown in panel (c), corresponding to the green line in panel (b).
}
\label{S1}
\end{figure*}

\begin{figure}[h!]
\centering
\includegraphics[width=0.45\textwidth]{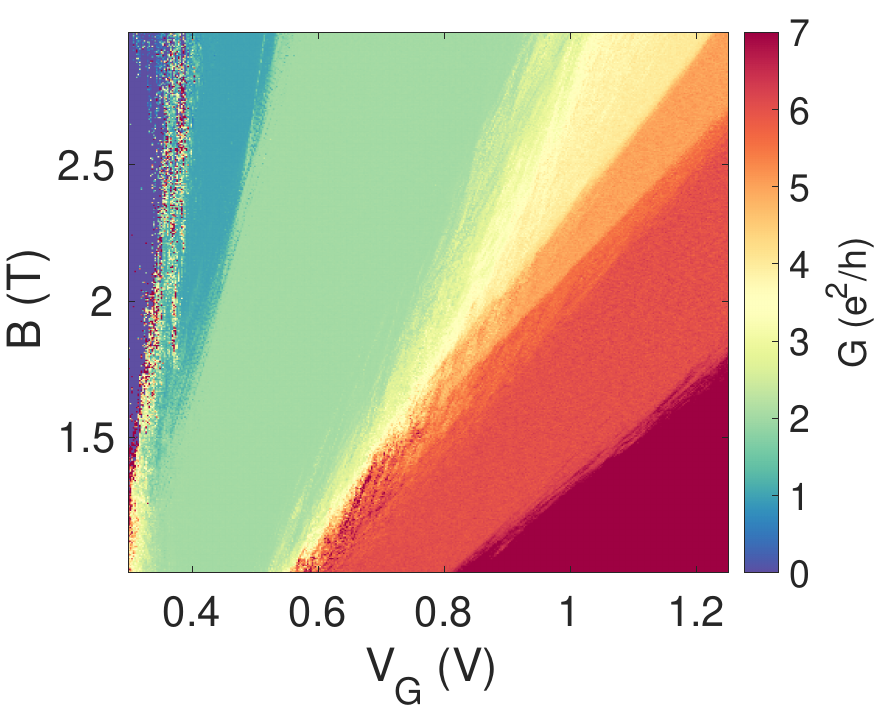}
\caption{Hall conductance of the region next to the $1\mu$m interface plotted vs. $B$ and $V_G$.}
\label{S2}
\end{figure}

We first verify that the downstream resistance fluctuations originate from the proximity effect of the superconductor instead of other spurious reasons (e.g. non-quantized QH bulk conductance). A careful examination of the signal has been presented in our former work~\cite{Zhao2020}. To be cautious, we still compare $R_d$ of the superconductor to that of a normal contact of the same interface length in Fig.~\ref{S1}. Indeed, the fluctuating $R_d$ is uniquely present only at the the superconducting interface. In addition, the level of QH quantization can be checked in the data presented in Fig.~\ref{S2}.

\section*{Data measured in additional samples}
\begin{figure}[h!]
\centering
\includegraphics[width=0.45\textwidth]{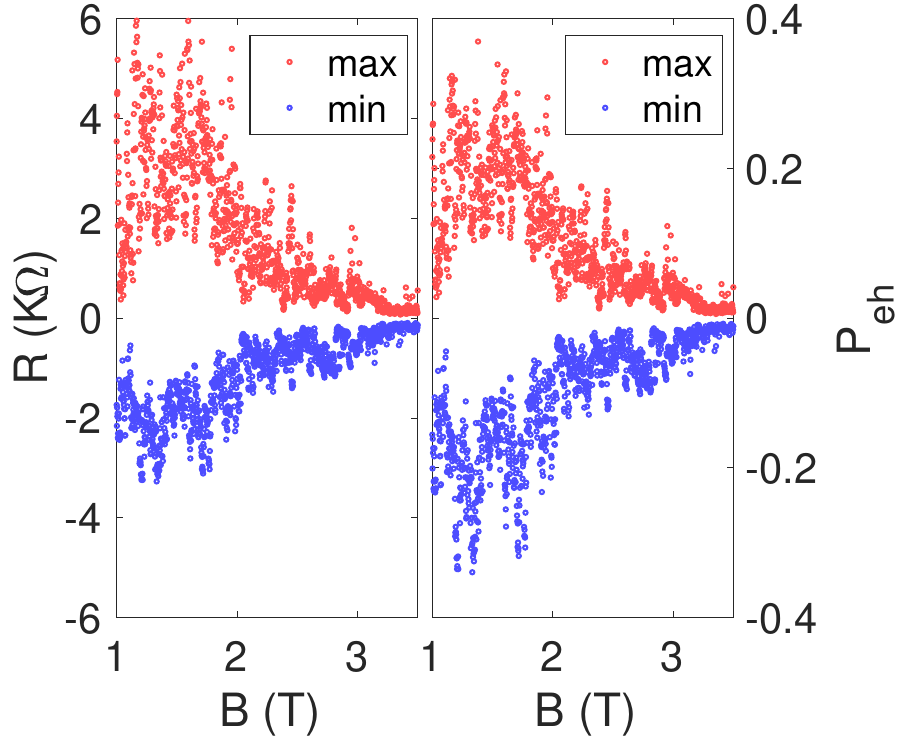}
\caption{Maximum and minimum of $R_d$ as a function of $B$ for another device. The length of the superconductor-graphene interface is 0.55 $\mu$m. }
\label{S3}
\end{figure}

In Fig.~\ref{S3}, we further illustrate the conversion from $R_d$ to $P_{eh}$ similar to that in Fig.~1c.

\begin{figure*}[h]
\centering
\includegraphics[width=0.8\textwidth]{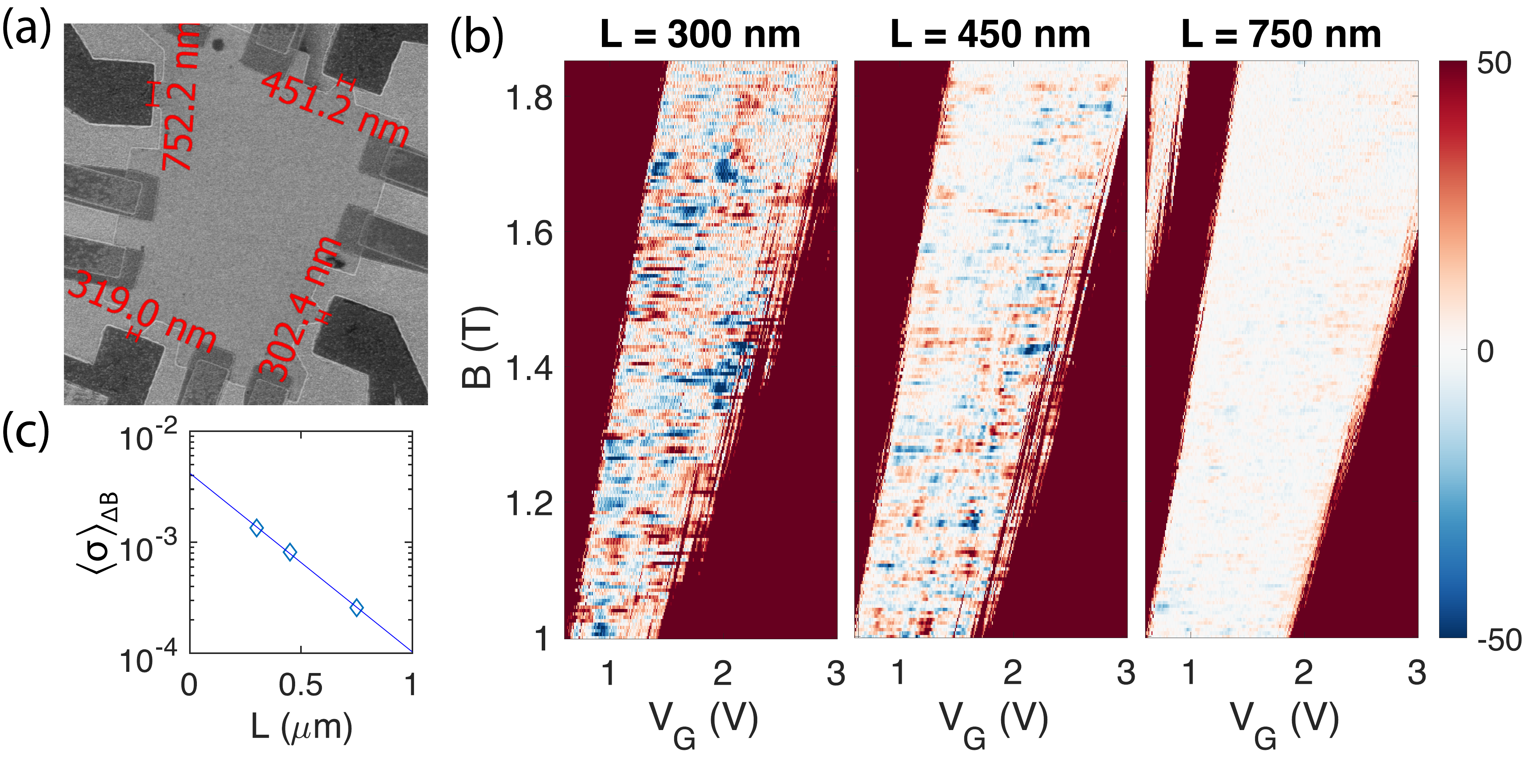}
\caption{(a) SEM image of another device with several contacts. (b) $R_d$ fans measured for different interface lengths. (c) $\left<\sigma\right>_{\Delta B}$ vs. $L$ at $B=$ 1.45 T.}
\label{S4}
\end{figure*}

To test the inferred relation $\left<\sigma\right>_{\Delta B}\propto\exp\left(-L/L_0(B)\right)$ with $L_0\propto1/B$, we further fabricated another device with 3 different $L$. In Fig.~\ref{S4} we plot $\left<\sigma\right>_{\Delta B}$ vs. $L$ at $T=$ 60 mK for $B$ in the $1.4-1.5$ T range (a 10 nA excitation is used for this measurement). Indeed, the dependence on $L$ clearly shows an exponential decay with $L_0\approx 270$ nm.

\section*{Uniform temperature dependence of $P_{eh}(V_G)$}

\begin{figure}[h!]
\centering
\includegraphics[width=0.45\textwidth]{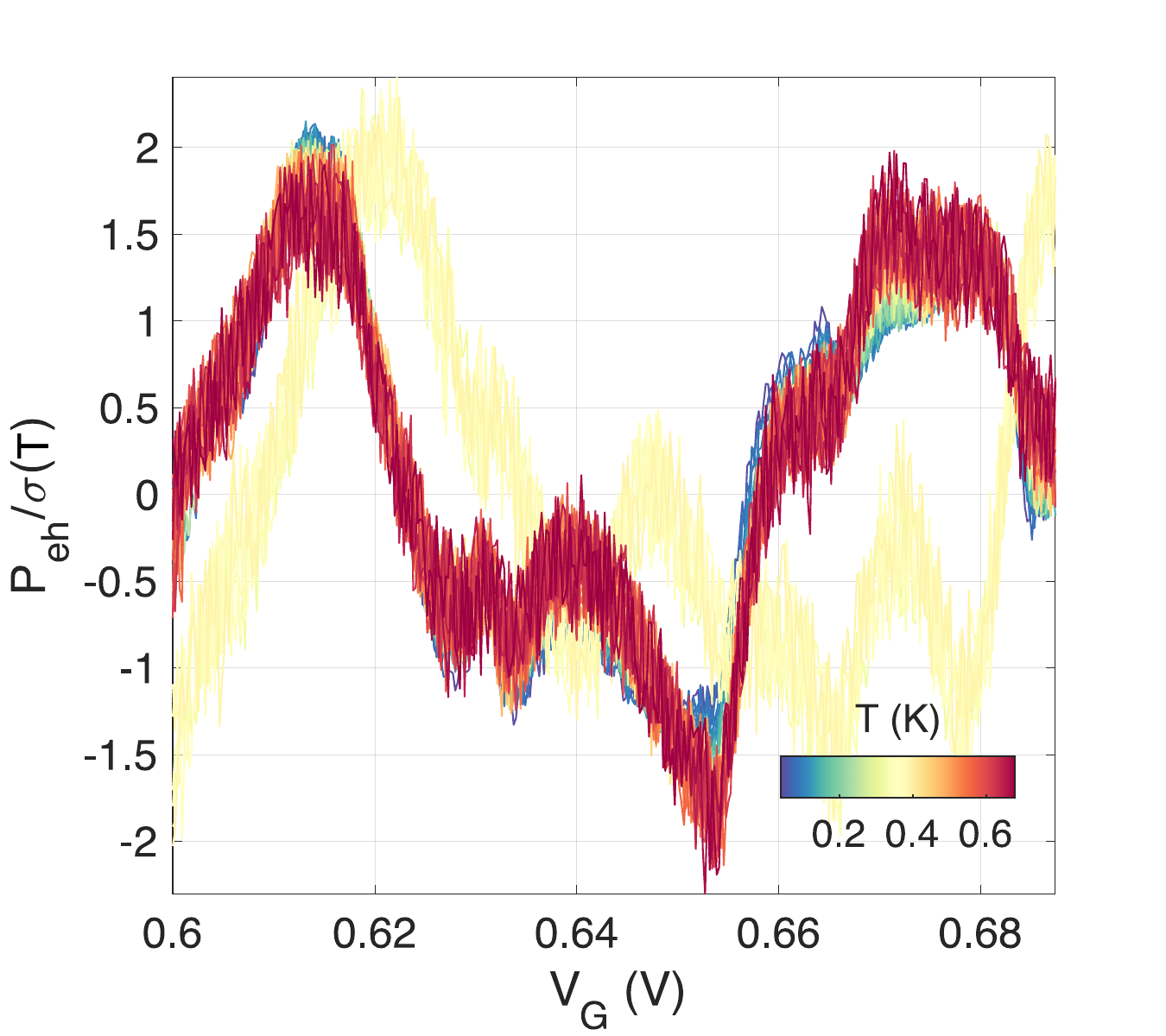}
\caption{$P_{eh}(V_G,T)$ curves normalized by their standard deviations $\sigma(T)$ for $T$ ranging from 40 mK to 670 mK. The corresponding unscaled $P_{eh}$ vs. $V_G$ curves are shown in Fig.~2a.}
\label{S7}
\end{figure}

In Fig.~2a, the $P_{eh}(V_G)$ curves gradually flatten with increasing temperature, without showing any noticeable broadening. In fact, the curves at different temperatures can be universally scaled by simply normalizing them by their standard deviation: $P_{eh}(V_G,T)/\sigma(T)$, where $\sigma(T)$ is calculated over the range of $V_G$ at the same $T$, Fig.~\ref{S7}. (This excludes the curves in yellow, that jump to a different pattern.) 
This observation allows us to explore $\sigma(T)$ vs. $T$ in place of the full $P_{eh}(V_G,T)$ in Fig.~2b. 

\section*{Variation of signal due to rearrangement of vortices}

\begin{figure}[h!]
\centering
\includegraphics[width=0.4\textwidth]{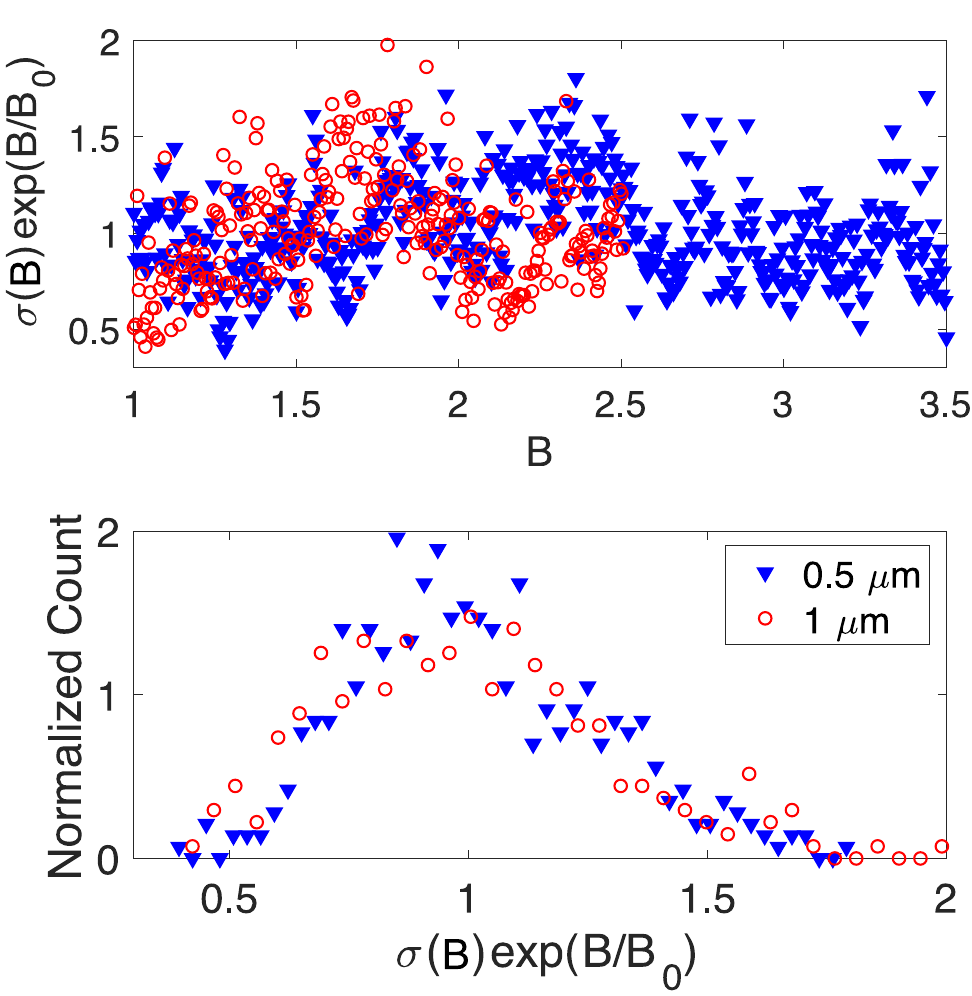}
\caption{$A(\{\textbf{r}_{vortex}\})=\sigma (B)\times\exp\left(\frac{B}{B_0}\right)$ plotted as a function of $B$ (top) and the corresponding histograms (bottom).}
\label{S6}
\end{figure}

Here we analyze the strong fluctuations of the signal, described by the prefactor $A(\{\textbf{r}_{vortex}\})$ in eq.~(1). At the lowest temperature, we can take $\sigma(B) = A(\{\textbf{r}_{vortex}\}) \times \exp\left(-\frac{B}{B_0}\right)$. Qualitatively, the vortex density increases with magnetic field, leading to the losses growing $\propto \exp\left(\frac{B}{B_0}\right)$. On top of this trend, the losses randomly change each few mT due to the vortex rearrangements. These variations are described by $A(\{\textbf{r}_{vortex}\})=\sigma(B) \times\exp\left(\frac{B}{B_0}\right)$, which is plotted in the top panel of Fig.~\ref{S6}. 
The quantity appears to be uniformly distributed in $B$. We plot its histogram in the bottom panel of Fig.~\ref{S6}. The distribution is broad and clearly asymmetric, with most of the counts ranging from 0.5 to 1.5 of the average value. 

\section*{Histograms of normalized $\sigma$}

\begin{figure*}[h!]
\centering
\includegraphics[width=0.7\textwidth]{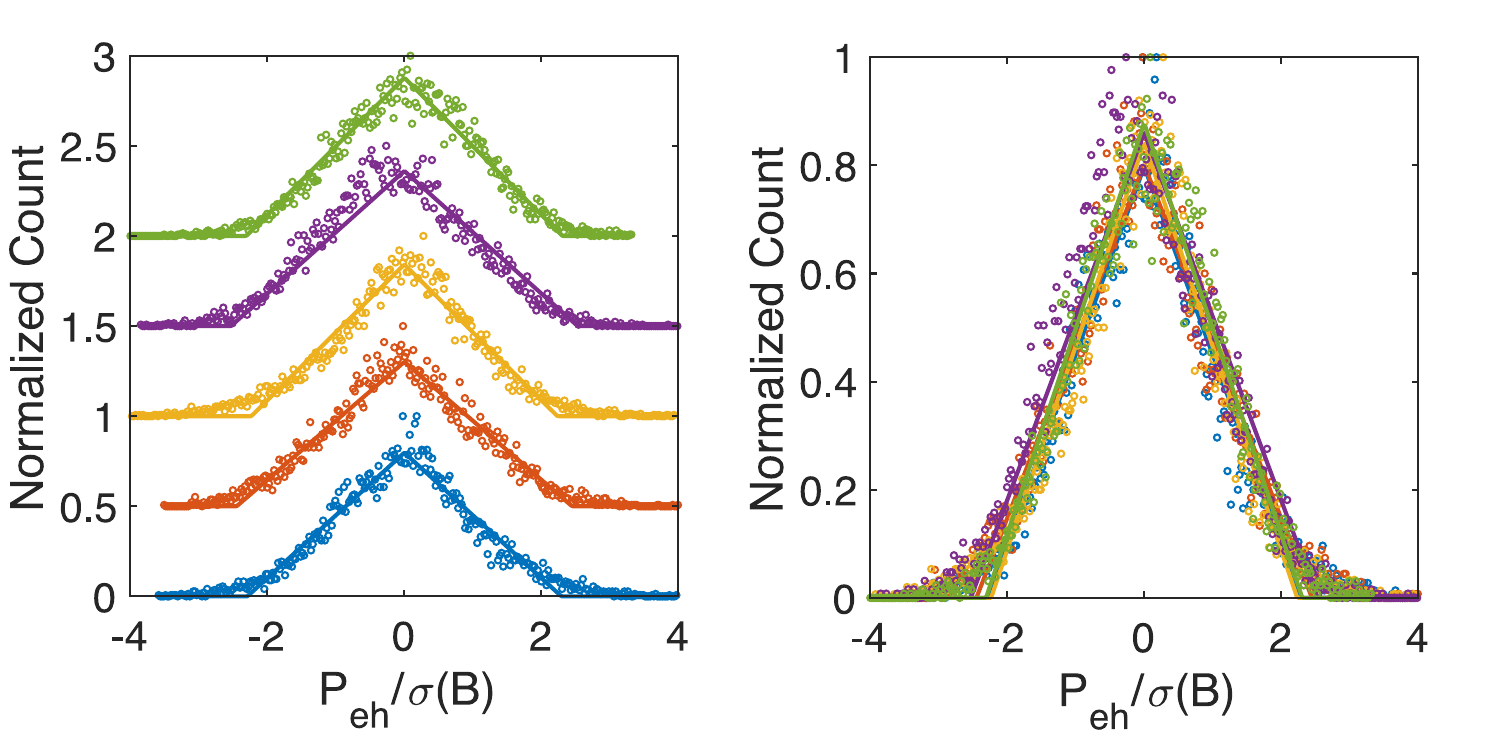}
\caption{Histograms of $P_{eh}(V_G,B)/\sigma(B)$ at different temperatures, plotted with an offset (left) and overlapping (right). We use the same color coding as in Fig.~4 of the main text. Normalization by the standard deviation $\sigma(B)$ allows one to remove the variation of losses due to the vortex rearrangement (as shown in Fig.~\ref{S6}). However, the procedure does not qualitatively change the histogram as compared to Fig.~4.}
\label{S5}
\end{figure*}

The distribution of $P_{eh}$ plotted in Fig.~4a is peaked around zero in contrast to the predicted uniform distribution in Ref.~\cite{Kurilovich_disorder_2022}. However, a uniform distribution can also be distorted if the distribution of vortex configurations $\{\textbf{r}_{vortex}\}$ produce a wide spread of $A\left(\{\textbf{r}_{vortex}\}\right)$ (see Fig.~\ref{S6}). To eliminate this effect, we normalize each $P_{eh}(V_G, B)$ curve by its standard deviation, $\sigma(B)$, calculated at the same $B$. We plot the histograms of the resulting quantity in Fig.~\ref{S5}. We find that the shape of the distribution remains roughly the same as in Fig.~4a. Therefore, the variation of losses with $B$ cannot explain the roughly triangular shape of the distribution. 

\section*{Toy model for the distribution of $P_{eh}$}

To explain the non-rectangular distribution in Fig.~4 and Fig.~\ref{S5}, we constructed a simple toy model, in which the local state of the CAES is represented by the $2 \times 2$ density matrix, $\rho$. The CAES propagation along the interface results in random trajectories on the Bloch sphere, which we simulate with the attached code. In the simulation, we assume a random variation of the phase of the order parameter along the interface~\cite{Kurilovich_disorder_2022}. We generate multiple realizations of the random superconducting phase to simulate the effect of the vortex rearrangements with $B$. For each realization, we assume that $V_G$ controls the coupling strength between the CAES and the superconductor. We vary this strength thereby generating different trajectories on the Bloch sphere. We take $P_{eh}=\rho_{11}-\rho_{22}$ at the final point of the interface, which yields the $P_{eh}(V_G)$ traces. We verify that without the losses or decoherence, the expected rectangular distribution of $P_{eh}$ is obtained, which corresponds to the uniform random coverage of the Bloch sphere (not shown).

\begin{figure*}[h!]
\centering
\includegraphics[width=0.8\textwidth]{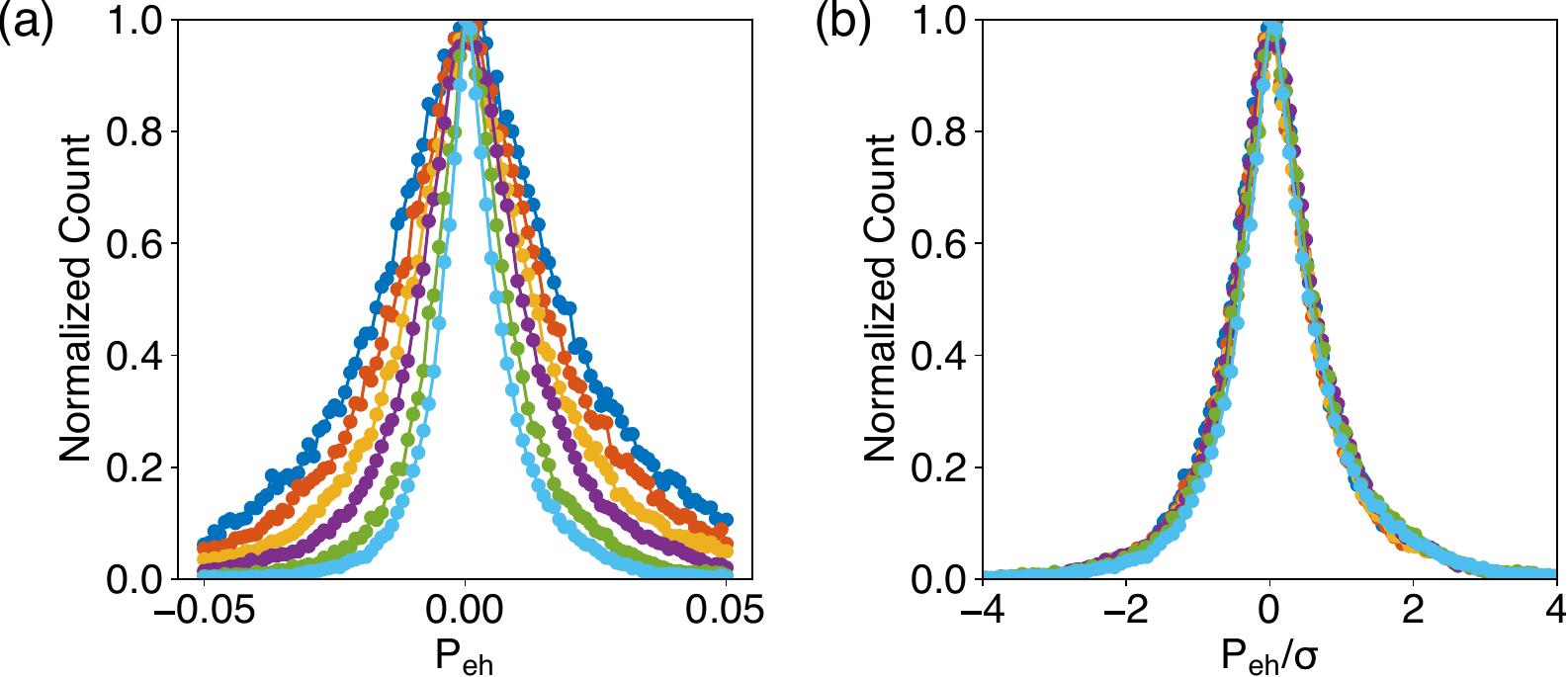}
\caption{(a) Simulated distributions of $P_{eh}$, as described in the text. In the widest curve (blue) only the losses are present, while progressively narrower curves add and increasing amount of decoherence. (b) The simulated distributions have very similar shapes, as could be revealed by rescaling the horizontal axis.}
\label{S8}
\end{figure*}

We first model the losses, which we assume to be due to tunneling into zero energy states, e.g. to the Caroli-de Gennes-Matricon states of the nearby vortices~\cite{Caroli1964}). We assume that the losses are localized at a few places along the interface corresponding  to the random locations of the closest vortices. At a given location along the interface, the loss is maximal at a certain position on the equator of the Bloch sphere, as determined by the local phase of the order parameter. The simulation produces a sharp distribution (blue curve in Fig.~\ref{S8}a), qualitatively similar to those found in the experiment. Analyzing the simulated trajectories, we find that majority of trajectories experience significant losses, resulting in the peak at zero $P_{eh}$. It is the outlier trajectories which happened to avoid heavy losses that contribute to the tails of the distribution.

We then introduce the dephasing of the standard $T_2$ type, which reduces coherence between the electron and hole components of the density matrix. Keeping the same randomly generated patterns of loss and the phase of the order parameter, we run the simulation for several values of the dephasing. Fig.~\ref{S8}a shows the resulting distributions, which appear qualitatively similar to Fig.~4. The shape of these distributions is also similar to the one obtained in the simulation with pure losses and no dephasing. The similarity of the simulated distributions allows us to rescale them as shown in Fig.~\ref{S8}b, similar to the rescaling of the experimental curves in Fig.~\ref{S5}b. 

Given the number of unknown parameters in our toy model, we do not pursue a qualitative agreement with the experiment. Instead, we believe the simulations illustrate the underlying reasons for the unusual shape of the observed histograms of $P_{eh}$. 



\end{document}